\documentclass{article}

\usepackage{arxiv}

\usepackage[utf8]{inputenc} 
\usepackage[T1]{fontenc}    
\usepackage{hyperref}       
\usepackage{url}            
\usepackage{booktabs}       
\usepackage{amsfonts}       
\usepackage{nicefrac}       
\usepackage{microtype}      
\usepackage{lipsum}
\usepackage{graphicx}
\graphicspath{ {./images/} }

\title{Four Guiding Principles for Modeling Causal Domain Knowledge: A Case Study on Brainstorming Approaches for Urban Blight Analysis}

\author{ \href{https://orcid.org/0000-0002-9527-5781}{\includegraphics[scale=0.02]{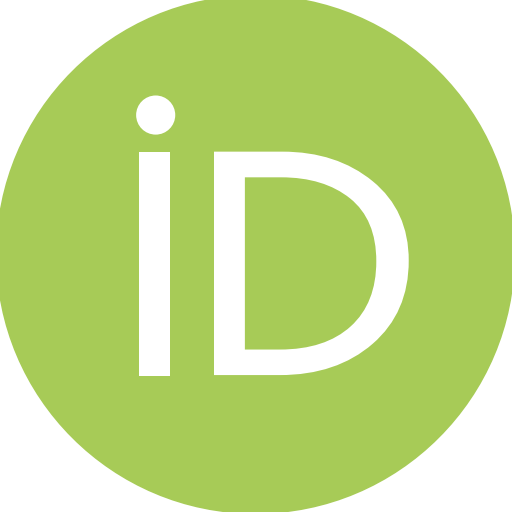}\hspace{1mm}Houssam Razouk}\thanks{Corresponding Author: \texttt{Houssam.razouk@student.tugraz.at} } \\
	Graz University of Technology, Graz, Austria\\
	\And
	\href{https://orcid.org/0000-0002-1204-0822}{\includegraphics[scale=0.02]{orcid.png}\hspace{1mm}Michael Leitner} \\
	Louisiana State University, Baton Rouge, United States\\
        \And
        \href{https://orcid.org/0000-0003-0202-6100}{\includegraphics[scale=0.02]{orcid.png}\hspace{1mm}Roman Kern} \\
	Graz University of Technology, Graz, Austria\\
        Know-Center GmbH, Graz, Austria \\
}

\begin{document}
\maketitle

\begin{abstract}
Urban blight is a problem of high interest for planning and policy making.
Researchers frequently propose theories about the relationships between urban blight indicators, focusing on relationships reflecting causality.
In this paper, we improve on the integration of domain knowledge in the analysis of urban blight by introducing four rules for effective modeling of causal domain knowledge.
The findings of this study reveal significant deviation from causal modeling guidelines by investigating cognitive maps developed for urban blight analysis. 
These findings provide valuable insights that will inform future work on urban blight, ultimately enhancing our understanding of urban blight complex interactions.
\end{abstract}


\maketitle

\section{Introduction}

Urban blight is defined as the deteriorating property conditions that have deleterious effects on the community in which the property is situated~\cite{beers2011quick}. 
Many scholars argued that blight might not be defined by how it looks, but it could be more determined by what it takes to reverse it~\cite{schilling2016basics}.
Therefore, urban blight is commonly estimated by a devising set of indicators.
Typically, theories about relationships between the urban blight indicators involve causality.
For example, the Broken Windows Theory~\cite{kelling1982broken}, suggests that urban blight contributes to increased crime in a neighborhood.
Generally, literature suggests that urban blight affects community welfare of the population in the affected neighborhoods ~\cite{LARSON201957}. 

In many cases, theories about relationships between the urban blight indicators (such as the Broken Windows Theory) are debated in the scientific community~\cite{rowan2017illusion, ansfield2020broken}. 
This controversy is mainly attributed to insufficient observations and the ethical concerns surrounding interventional experiments. 
Also, it is important to keep in mind that correlation does not imply causation. 
Causal statements derived from observational data are only valid under specific circumstances and strong assumptions. 
Additionally, the resources and technologies limitations impede the data collection at a scale.
Consequently, the available data sets are typically small and more likely to be biased.
These biases severely affect the data analytics methods results reliability and robustness as well as the accuracy of the drawn conclusions.

At the same time, researchers in the field of causal data science demonstrated that data analytics methods benefit from integrating domain knowledge.
Integrating domain knowledge showed its effectiveness in correcting for data biases and achieving more reliable results~\cite{pearl2018theoretical,bareinboim2016causal}.
One great example of the effective use of domain knowledge is addressing the Simpson's paradox~\cite{pearl2011simpson}.
In this example, Pearl~\cite{pearl2011simpson} illustrates that the paradox resolution relies on the fact that causality is governed by its own logic and cannot be resolved with observational data alone.
This logic is derived mainly from domain knowledge which is acquired by the subject matter experts by observations and intervening with the environment under study.
To this end, in the case of urban blight analysis, domain knowledge is critical.

 To gather and represent subject matter experts' domain knowledge regarding a particular topic, brainstorming approaches, such as cognitive mapping, are commonly utilized~\cite{eden2004cognitive,rodrigues2022artificial,vieira2022measuring}. 
In the case of urban blight, these approaches have been applied extensively~\cite{pinto2023analyzing,assunccao2020rethinking}.
However, brainstorming approaches, e.g. cognitive mapping, also have several key limitations. 
For instance, if a different team of experts was selected, the resulting model could differ significantly~\cite{pinto2023analyzing}. 
Furthermore, by reviewing the generated domain knowledge model, the criteria used in the modeling of domain knowledge, with regards to urban blight using cognitive mapping, may not be clear and could overlook some potential pitfalls common in causal domain knowledge modeling.
These existing limitations pose significant challenges when effectively modeling urban blight domain knowledge. 
To this end, the following research questions are raised:

\begin{itemize}

    \item What are the prevalent challenges and pitfalls associated with modeling causal domain knowledge using cognitive mapping and other brainstorming approaches in the context of urban blight analysis?

    \item In what ways can the recent advances in the field of causal data science improve the modeling guidelines of causal domain knowledge acquired through brainstorming approaches such as cognitive mapping?

\end{itemize}

These research questions are addressed by reviewing advances in causal data science and emphasizing the necessity of establishing close collaboration between subject matter experts and experts in the field of causality. 
As such, the overall contribution of this research is to provide guidance for the causal modeling step in effectively integrating domain knowledge for future research on urban blight.
Specifically, the contribution of this paper is as follows:

\begin{itemize}

    \item Introducing the field of causal data sciences to scientific and practical communities interested in urban blight analysis 
    \item Contribute to the generation of more reliable urban blight analysis by identifying and highlighting common pitfalls associated with the use of brainstorming approaches, such as cognitive mapping, for causal domain knowledge models.
    \item Enhancement of modeling guidelines for causal domain knowledge acquired through brainstorming approaches building upon recent advances in field of causal data science. 
    
\end{itemize}

\section{Background and related work}

A comprehensive review of relevant literature related to urban blight is provided in Section~\ref{Introduction to urban blight}.
Additionally, the use of cognitive mapping in the context of urban blight analysis is summarized in Section~\ref{Cognitive mapping based approaches for urban blight analysis}.
Finally, a short introduction to causal data science is detailed in Section~\ref{Introduction to causal data science}.

\subsection{Introduction to urban blight}
\label{Introduction to urban blight}

In the 20th century, many countries experienced both urbanization and suburbanization. 
Urbanization is the process of people moving from rural areas to urban areas~\cite{curci2017flight}.
Urbanization can bring increased income, improved living standards, and access to better education and healthcare, but it also has numerous negative consequences.
For example, crime and violence are mostly common and severe in urban areas~\cite{world2010hidden}. 
In contrast to urbanization, suburbanization is the movement of population from urban city centers to less densely populated areas, such as suburbs and rural towns~\cite{curci2017flight}.
As people move from urban city centers to less densely populated areas during suburbanization, there can be a decline in population in the urban cities, which can lead to an increase in the number of vacant properties~\cite{sugrue2014origins}. 
In extreme cases, the decline in population in an urban area can lead to the abandonment of properties.
For instance, in Detroit (USA), nearly 90,000 vacant properties were reported in 2013~\cite{sugrue2014origins}.
Consequently, urban areas can become predominantly populated by groups, who may face various social and economic challenges such as poverty, unemployment, and lack of access to education and healthcare~\cite{de2006residential}.
These social and economical issues together with vacant properties and brownfields (i.e. previously developed land that is not currently in use) can contribute to phenomena known as urban blight~\cite{de2006residential}.

Urban blight is a term that has multiple interpretations and evaluations due to its subjective nature and the variety of disciplines it encompasses. 
Thus, there is no universally accepted definition for this phenomenon~\cite{pinto2023analyzing}. 
Further, Schilling and Pinz{\'o}n~\cite{schilling2016basics} argued that the definition of urban blight should not be based solely on its appearance, but rather on the actions and efforts required to reverse it.
Thus, despite the ambiguity of the term, many scholars associate urban blight with high crime rates~\cite{zhang2014addressing}, vacant residential and commercial properties, reduced housing quality, a decline in local services~\cite{kondo2018blight}, and overall deterioration of urban areas~\cite{rafiee2013urban}.

In some cases, urban blight, as a widespread phenomenon, is considered a potential cause of other drastic consequences such as drug trafficking, prostitution, and violence~\cite{pinto2023analyzing}. 
These cases align with the work of Kelling et al.~\cite{kelling1982broken} who viewed urban blight through the "Broken Windows Theory".
Namely, urban blight can be identified by the presence of vacant properties and abandoned facilities that reflect the physical, social, and economic conditions of an area.
Moreover, urban blight can also be indicated by a lack of adequate housing, whereas in the mid twentieth century the concept of urban blight included the depreciation of property and a decreased demand for services~\cite{maghelal2014assessing}. 
More recently, urban blight is also characterized by urban poverty, ghettos, slums, high population density (overcrowded areas), high crime rates, and a concentration of minority households in blighted areas~\cite{maghelal2014assessing}. 

Much of the research on urban blight differentiates between physical and social indicators. 
For instance, the study on blight done by Maghelal et al.~\cite{maghelal2014assessing} mapped physical indicators in the City of Dallas into seven subcategories: abandoned, vacant residential, vacant commercial, mortgage foreclosed, tax foreclosed property, tax delinquent, and demolished. 
Next, Maghelal et al.~\cite{maghelal2014assessing} examined the socioeconomic blight indicators and mapped them into the following categories: poverty, unemployment, ethnicity, race, renter household, population, and single parent household. 

To account for both physical and social indicators, recent research on urban blight generally utilizes mixed method approaches, by combining both qualitative and quantitative factors.
This is illustrated by Rafiee and Mahesh~\cite{rafiee2013urban} where the authors leverage a triangulation approach to assess the main causes of urban blight in the historical centre of Shiraz city. 
Another example is presented by Fabiyi~\cite{fabiyi2011analysis} where the author has analyzed the spatial and temporal pattern of urban blight in metropolitan Ibadan (Nigeria) through the data obtained from satellite images, questionnaire, and indicators identified by experts.
Furthermore, Hosseini and Hajilou~\cite{hosseini2019drivers} evaluated urban blight in Laleh-Zar neighborhood of Tehran by using literature review, statistical methods and questionnaires filled in by random samples of residents in the area.
Moreover, De Tuya et al.~\cite{de2017information} aimed to identify data requirements to create value in the context of urban blight by including stakeholders feedback through workshops, meetings and focus groups. 

In summary, most studies on urban blight have similar methods, leading to similar limitations.
These include a focus on a single neighborhood or location and reliance on the subjective opinions of a small group of experts. 
We believe that some of these limitations stem from the limited availability of domain knowledge and the difficulty of obtaining and modeling it.

\subsection{Cognitive mapping based approaches for urban blight analysis}
\label{Cognitive mapping based approaches for urban blight analysis}

To study the causes of urban blight, Pinto et al.~\cite{pinto2023analyzing} devised a multiple criteria decision analysis (MCDA) approach. 
Namely, Pinto et al.~\cite{pinto2023analyzing}  leverage a cognitive mapping based method to identify, select, categorize and group potential causes of urban blight. 
Next, to perform quantitative analysis, the authors devise the decision making trial and evaluation laboratory (DEMATEL) method. 
Consequently, several limitations of this study are noted by the authors.
These limitations include: 
(i) If a different group of subject matter experts are involved in the study, different causes of urban blight might have been identified;
(ii) The limited number of panel members and the constrained study area could affect the representativeness of the study.

Also, the study by Costa et al.~\cite{costa2021intervention} presents a simplified approach to tackling the blight problem by employing unique techniques for structuring complex decision problems, known as problem structuring methods. 
These methods are rooted in the principles of multiple-criteria decision analysis (MCDA) and incorporate techniques for conducting interviewees with domain experts such as decision conferencing and cognitive mapping. 
Furthermore, the study utilizes the decision-making trial and evaluation laboratory (DEMATEL) technique in a subsequent phase to facilitate the development of a more realistic and transparent analysis model for blight intervention strategies.
In Costa et al.'s study, the interviewees highlighted certain limitations, particularly underscoring the challenge posed by the relatively small size of the expert panel.
The interviewees expressed concern that despite efforts to ensure diversity within the small sample, the limited number of participants may hinder the extrapolation of results to other contexts, regardless of their diversity.

Similarly to Pinto et al.'s research, Ferreira et al.~\cite{ferreira2022urban} conducted a study on cause and effect relationships of urban blight prevention strategies. 
Ferreira et al.'s research methodology involve cognitive mapping and multiple criteria decision analysis (MCDA). 
Moreover, Ferreira et al.~\cite{ferreira2022urban} opted for the MA-DEMATEL method aiming for more informed decision making.
For readers who are interested in the MA-DEMATEL method, the method extends DEMATEL by incorporating moving averages (MAs).
In Ferreira et al.'s study, cognitive maps are generated devising the strategic options development and analysis (SODA) approach. 
Various prevention initiatives are identified and organized into clusters (areas of concern) by the subject matter experts.
The subject matter experts that were selected for the decision maker panel were urban architects, civil engineers, representatives of homeowner associations, and senior city planners within the Lisbon city council.
Although Ferreira et al. opted for the novel MA-DEMATEL approach to eliminate the limitations from previous studies on urban blight, concerns were raised by the subject matter experts.
These concerns include: (i) simplified causal map, (ii) the model building process is context dependent and includes subjective components, and (iii) limitations related to the comprehensibility and  representativeness of the resulted cognitive map.
These limitations are similar to the ones provided in a study done by Pinto et al.

Further, to create a conceptual model of the causes of blight, Lousada et al.~\cite{lousada2021sociotechnical}  used fuzzy cognitive mapping (FCMs) which are graphical representations of causal reasoning. 
Specifically, to study the nonlinear behavior of complex systems over time, authors included a system dynamics approach (SD). 
Here, an aggregation of experts' opinions are presented to study the cause and effect relationships between urban blight causes.
In Lousada et al.'s  study, the expert panel is composed of 7 members with high level expertise in urban blight.  
The listed method limitations in Lousada et al.'s  study include:
(i) difficulty of finding experts specialized in urban blight; 
(ii) time consuming;
(iii) differences in opinion among experts, (i.e. subjectivity);
(iv) difficulty in consolidation of the results.
In addition, limitations similar to the ones mentioned by  Pinto et al.~\cite{pinto2023analyzing} are noted. 

In general, brainstorming approaches, such as cognitive mapping based approaches, allow for the integration of multiple perspectives and sources of information, which can lead to a more comprehensive understanding of the problem.
However, these approaches have several key limitations. 
One of the main limitations is that  brainstorming approaches are time consuming.
Additionally, these methods are often vulnerable to biases introduced by the moderator.
Furthermore,  brainstorming approaches have limited coverage, especially when researchers have access to only a limited number of subject matter experts. 
Finally, the criteria used in modeling causal domain knowledge regarding urban blight using cognitive mapping may not be clear, and this might lead to overlooking of some common pitfalls in causal domain knowledge modeling.

\subsection{Introduction to causal data science}
\label{Introduction to causal data science}

Causal data science extends traditional data science by explicitly considering the underlying data generation process, which can be used to improve the model prediction robustness and reliability.
As such, causal data science is gaining more interest in parallel to the rapid development of data driven approaches, such as machine learning algorithms, and the era of big data~\cite{Guo_2020}.
In machine learning, the prediction quality is highly dependent on the quality of the validity of the assumptions made during the data analytic process.
Greenland summarized the most common causal data science frameworks in his work~\cite{greenland2002overview}.
Nowadays causal frameworks includes, but is not limited to, graphical causal models (causal diagrams), potential outcome models, structural equations models that further inspired structural causal model, and sufficient component causal models.
Causal diagrams, potential outcome models, and structural causal models are all tools that can be used to analyze cause and effect in a study population. 
Causal diagrams are useful for providing a clear, easy to understand visual representation of the qualitative assumptions behind a causal analysis.
Potential outcome and structural equations models, on the other hand, can be used to articulate more detailed quantitative estimates about how different units in the population may respond to different factors. 
Sufficient component causal models are distinct from these other models in that they depict more complex qualitative assumptions about the specific causal mechanisms by which different factors may be causing different effects within individual units of the population~\cite{greenland2002overview}.
In summary, all these causal models can be used to understand the relationships between causes and effects, however, different types of causal models provide different information granularity levels and focus on different aspects of the study population.

The primary focus of this research is to enhance the modeling of causal domain knowledge as it pertains to urban blight.
To achieve this objective, the research method leverages the causal diagrams framework.
Causal diagrams refer to a set of variables and the direct cause effect relationships that exist between these variables. 
In this framework, the variables are depicted as nodes or vertices, while the cause effect relationships are depicted as directed edges or arcs. 
Causal diagrams have been demonstrated to be an effective tool in helping scholars gain a deeper understanding of complex problems.
For instance, causal diagrams have been particularly helpful in scenarios like Simpson’s paradox, where the relationship between different variables may not be immediately obvious.
Simpson's paradox is a statistical phenomenon that occurs when a trend appears in different groups of data but disappears or reverses when these groups are combined.
According to Pearl~\cite{pearl2011simpson}, the key takeaway from Simpson's paradox is that the results obtained from the stratified data set are more reliable than those obtained from the aggregated data set if the stratification variable (which defines the groups) introduces confounding bias.
However, if the stratification variable does not introduce confounding bias, then the correct results are found using the aggregated data set. 
This underscores the importance of considering potential sources of bias when interpreting research results, and highlights that the reliability of the findings depends on the specific context. 
At the same time, it is worth noting that the results obtained from either the stratified data set or the aggregated data set do not necessarily mean that one set is correct and the other is incorrect, but rather that one set is more reliable than the other.

Causal diagrams are a powerful tool for causal modeling that can be used to improve the robustness and reliability of predictions made by a model. 
However, using causal diagrams for causal modeling in real-world scenarios can be challenging due to potential pitfalls that can easily be overlooked~\cite{suzuki2020causal}. 
To address this issue, Suzuki et al.~\cite{suzuki2020causal} have provided a set of guidelines for scholars who are using causal diagrams, particularly causal directed acyclic graphs (DAGs). 
These guidelines include representing all causal relationships present in the population, regardless of how frequently these causal relationships occur.
This is important because it ensures that the diagram is as comprehensive as possible, capturing all potential causal relationships.
Another important tip is to represent the nodes in the causal diagram as the causal variable and not its realized value. 
This approach ensures that the diagram represents the underlying causal mechanisms, rather than just the observable values of the variables.

To ensure the validity of causal models, it is crucial to consider the principles of transitivity and proportionality in causation, as outlined in Neil McDonnell's work~\cite{mcdonnell2018transitivity}. Transitivity is a logical principle that states that if A causes B, and B causes C, then A must also cause C. This principle is often used as a criterion to check whether the relationship between variables is consistent. When transitivity is violated, it may indicate that the causal model needs to be revised. 
As an example, if we assume that "pollution causes respiratory disease" (A causes B) and "respiratory disease causes premature death" (B causes C), it must be true that "pollution causes premature death" (A causes C) if transitivity holds. By following this principle, scholars can ensure the validity of causal models and make accurate predictions regarding causal relationships between variables.

At the same time, the concept of modularity, as it applies to causal diagrams, has been discussed by Kuorikoski~\cite{Kuorikoski}.
Modularity is important because it allows different parts of a causal diagram to be separated and analyzed independently.
Kuorikoski~\cite{Kuorikoski} distinguishes between three types of modularity in causal diagrams.
Firstly, there is causal model modularity in variables. 
This refers to the ability to surgically intervene on a variable (i.e. assign a value to the variable) in the causal model without affecting other variables in the model.
Secondly, there is variable mechanism\footnote{The causal effect mechanism indicates the way the treatment affect the outcome} modularity.
This indicates the independence between the variable (or at least a range of the variable) and the causal effect mechanism. 
This ensures the smoothness of the outcome, meaning that the causal mechanism is independent from the variable values.
Thiredly, there is causal model modularity in parameters\footnote{parameters are the variables which are assumed to be constant when developing the causal model for a certain context}. 
This refers to the ability to intervene on a parameter in the causal model and affect only one causal effect relation in the model.
Overall, modularity allows for a more detailed and nuanced analysis of causal relationships in a system, by breaking down the system into its constituent parts and analyzing them independently.

In addition, the concept of sufficient structure causal diagrams, introduced by VanderWeele and Robins~\cite{VanderWeele_2009}, is a way to link different causal modeling frameworks (including causal diagrams) and provide a more detailed understanding of underlying causal mechanisms and new causal relationships. 
This approach involves adding  artificial nodes to the original causal diagrams, which represent sufficient causes.
By conditioning on a certain strata of the outcome in a binary setting, this allows for the discovery of conditional independence between the parents (the causes).
Overall, the concept of sufficient structure causal diagrams provides a more comprehensive approach to causal modeling, allowing for the discovery of new causal relationships and a better understanding of the underlying causal mechanisms. 
It serves as a useful tool for researchers in various fields who are seeking to gain a more detailed understanding of causal relationships in complex systems.

More recently, Zhang et al.~\cite{zhang2022causal} have  differentiated between explicit variable and generic variable. An explicit variable can be viewed as a specialized version of a generic variable, as it pertains to a specific unit, sample, or individual.
While a generic variable, such as "X", is applicable across a range of instances, an explicit variable, such as "(Xi)" is tailored to a specific unit i and its corresponding attribute.
By doing so, ~\cite{zhang2022causal} were able to propose definitions of the interaction model and the isolated interaction model.
These definitions are necessary to get more reliable results in settings where the IID (Independent Identical Distribution) assumption does not hold. 

In summary, the field of causal data science, which encompasses causal domain knowledge modeling, is constantly evolving. 
This is an exciting area of research that has the potential to transform the way we approach data analysis and decision-making. 
By leveraging causal modeling techniques and domain knowledge, a deeper understanding of complex phenomena can be gained and  causal relationships that may not be apparent using traditional statistical methods can be identified.
Ultimately, this has the potential to lead to more effective policies and interventions that can help improve the lives of individuals and communities.

\section{Method}

The purpose of this research is to enhance the modeling of causal domain knowledge commonly acquired through brainstorming approaches such as the cognitive mapping approach.
Specifically, causal graphical models framework represented by causal diagrams have been selected because they are useful for providing a clear, easy to understand visual representation of the qualitative assumptions behind a causal analysis~\cite{greenland2002overview}.
To this end, this research recommends a set of four rules to guide scholars and avoid common errors in causal modeling of domain knowledge.
The four rules are focused on identifying causal variables (Section~\ref{CMR1}), identifying artificial nodes (Section~\ref{CMR2}), identifying causal relations between variables (Section~\ref{CMR3}), and checking the encoding criterion for causal variables (Section~\ref{CMR4}).
By following these rules, scholars can improve their causal modeling and minimize errors.

\subsection{Nodes in the causal diagram represents causal variables not its realized values}
\label{CMR1}

In brainstorming approaches, such as cognitive mapping, subject matter experts are often asked to identify possible causes or effects of a given concept, such as urban blight. 
These causes are, typically, articulated as text entities that describe the value of the causal variable. 
Naturally, these text entities can describe different values of the same causal variable.
In some cases, text entities that describe the same value of a causal variable can be articulated differently and included multiple times under different naming convection.
However, if these text entities are represented as independent nodes in the causal diagram, these text entities can negatively impact the modularity of the proposed causal model. 
Specifically, modularity in variables can be compromised. 
Hence, the resulting causal model will suffer from biases because it does not model the relationship between these related text entities. 
Furthermore, the resulting causal model will not support interventions on these text entities  because intervening in one of these related text entities  would affect other related text entities.

\begin{figure}
  \centering
  \includegraphics[width=0.3\textwidth]{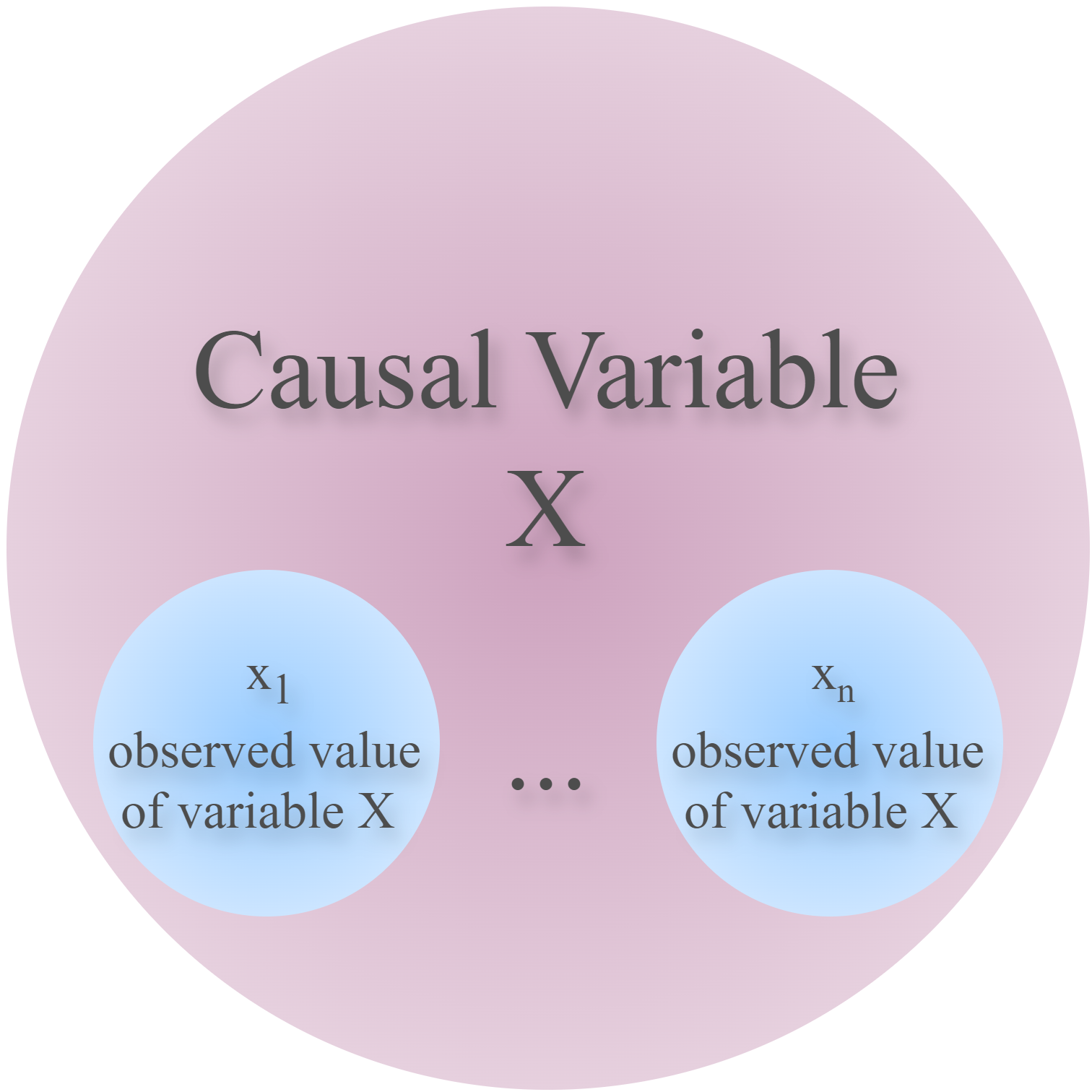}
  \caption{\textbf{Grouping related text entities under their corresponding causal variable.} The text entities $x_1$ to $x_n$ are articulated by subject matter experts. These text entities describe different values of the causal variable $X$. The relationships between these text entities can be identified as either mutually exclusive, as $X$ cannot take multiple values simultaneously, or as describing the same value of the causal variable.}
  \label{fig:Causal modeling-Rule_1}
\end{figure}

To address this, practitioners are highly encouraged to consider treating related text entities as a group and creating a single node to represent the underlying causal variable. 
To identify such related text entities, it is recommended to go through all the identified text entities and group those that represent different values of the same causal variable. 
Figure~\ref{fig:Causal modeling-Rule_1} provides a visual demonstration of an examples of grouping of text entities, denoted as $x_1$,...,$x_n$, which describe different values of the same causal variable, denoted as $X$.
The relation between $x_1$ and $x_n$ could be that $x_n$ is the same as $x_1$ or if $x_n$ is observed then $x_1$ cannot be realized due the fact that $X$ cannot take two values simultaneously.
This rule improves on the modularity of the model and provide a more accurate representation of the causal relationships between variables. 
Additionally, this approach enable practitioners to identify and test interventions that target specific causal variable within the model, without impacting the causal variables. 
This recommendation is highlighted as one of the tips provided by Suzuki et al.~\cite{suzuki2020causal} and  is in line with the concept of modularity in variables presented by Kuorikoski~\cite{Kuorikoski}.


\subsection{Interaction entities, that give information about more than one causal variable, are the result of their interaction and should be modeled as artificial nodes}
\label{CMR2}

\begin{figure}
  \centering
  \includegraphics[width=0.6\textwidth]{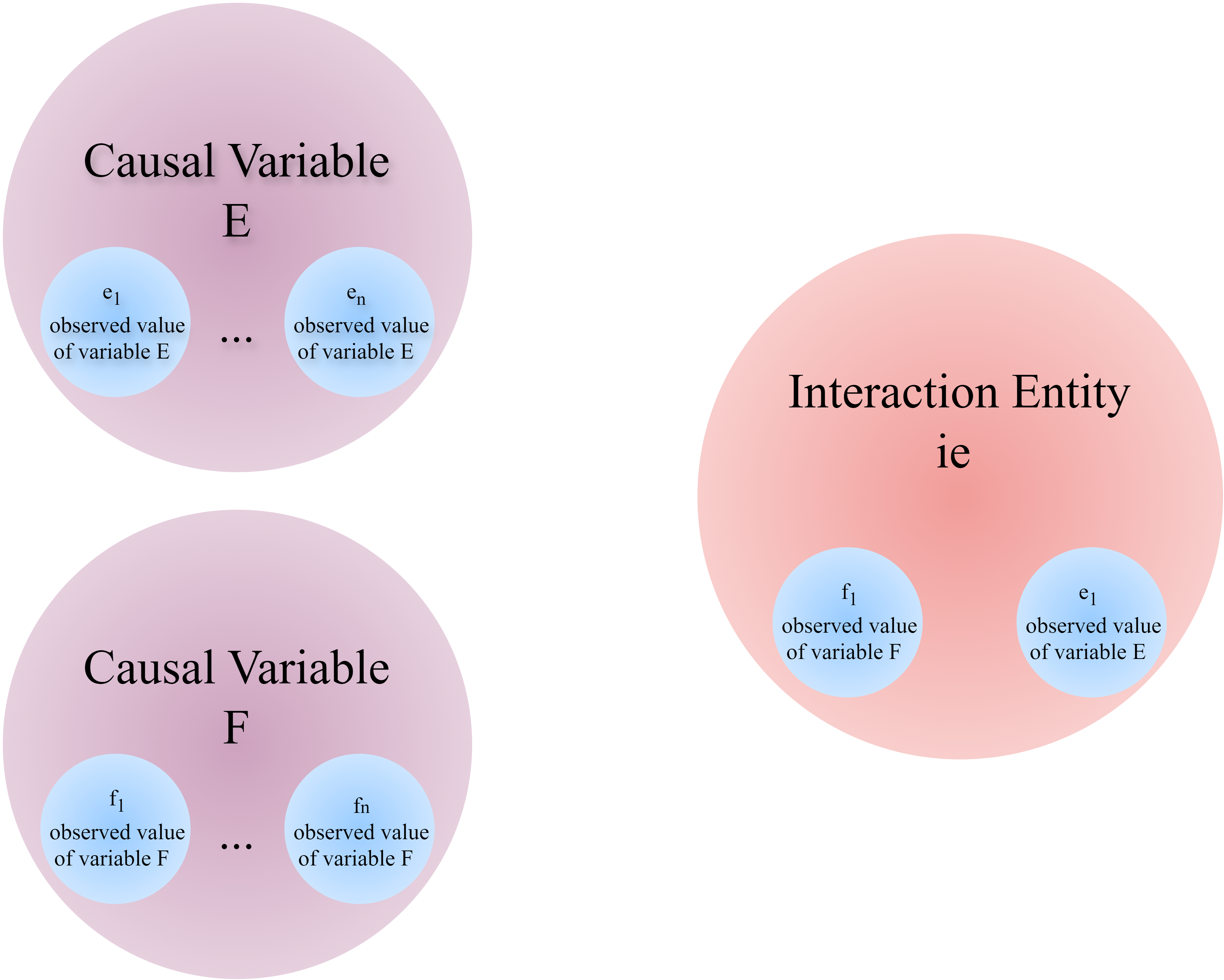}
  \caption{\textbf{Representing interaction entities as artificial nodes.} Interaction entities are text entities which describe an interaction between two or more causal variables. These text entities should be represented as artificial nodes in the causal diagram. The artificial nodes encompass the corresponding observed values of the causal variables it represents. The artificial node can be linked to causal variables; however the type of this relation should differ from other causal relations in diagrams.}
  \label{fig:Causal modeling-Rule_2}
\end{figure}

Brainstorming approaches for capturing causal domain knowledge, such as cognitive mapping,  relies on human subject matter experts in articulating the causes and effects of a certain  concept.
Thus, the causes and effects are articulated as text entities. 
In some cases, an text entity can describe an interaction between two or more causal variables. 
For instance, an interaction entity denoted as $ie$ describes an interaction between causal variable $F$ and variable $E$. 
Specifically, the interaction entity $ie$ corresponds to the simultaneous occurrence of observed value $f_1$ for causal variable $F$ and observed value $e_1$ for variable $E$.
It is important to note that such text entities should not be included in either causal variable $F$ or $E$ to maintain the modularity of these variables. 
Instead, representing the entity $ie$ in the causal diagram as an artificial node is recommended.
This artificial node should encompass the corresponding observed values it represents; in this example, $f_1$ and $e_1$.
Furthermore, the artificial node can be linked to the causal variables that encompass the observed values of this interaction entity. 
This approach aligns with the technique proposed by VanderWeele and Robins~\cite{VanderWeele_2009} for modeling interactions between two or more causal variables in causal diagrams. 
Figure~\ref{fig:Causal modeling-Rule_2} depicts an example of representing an text entity describing an interaction between two causal variables.


\subsection{Presence of causal relations between nodes in causal diagrams corresponds to the presence of at least one causal relation between two text entities}
\label{CMR3}

\begin{figure}
  \centering
  \includegraphics[width=0.6\textwidth]{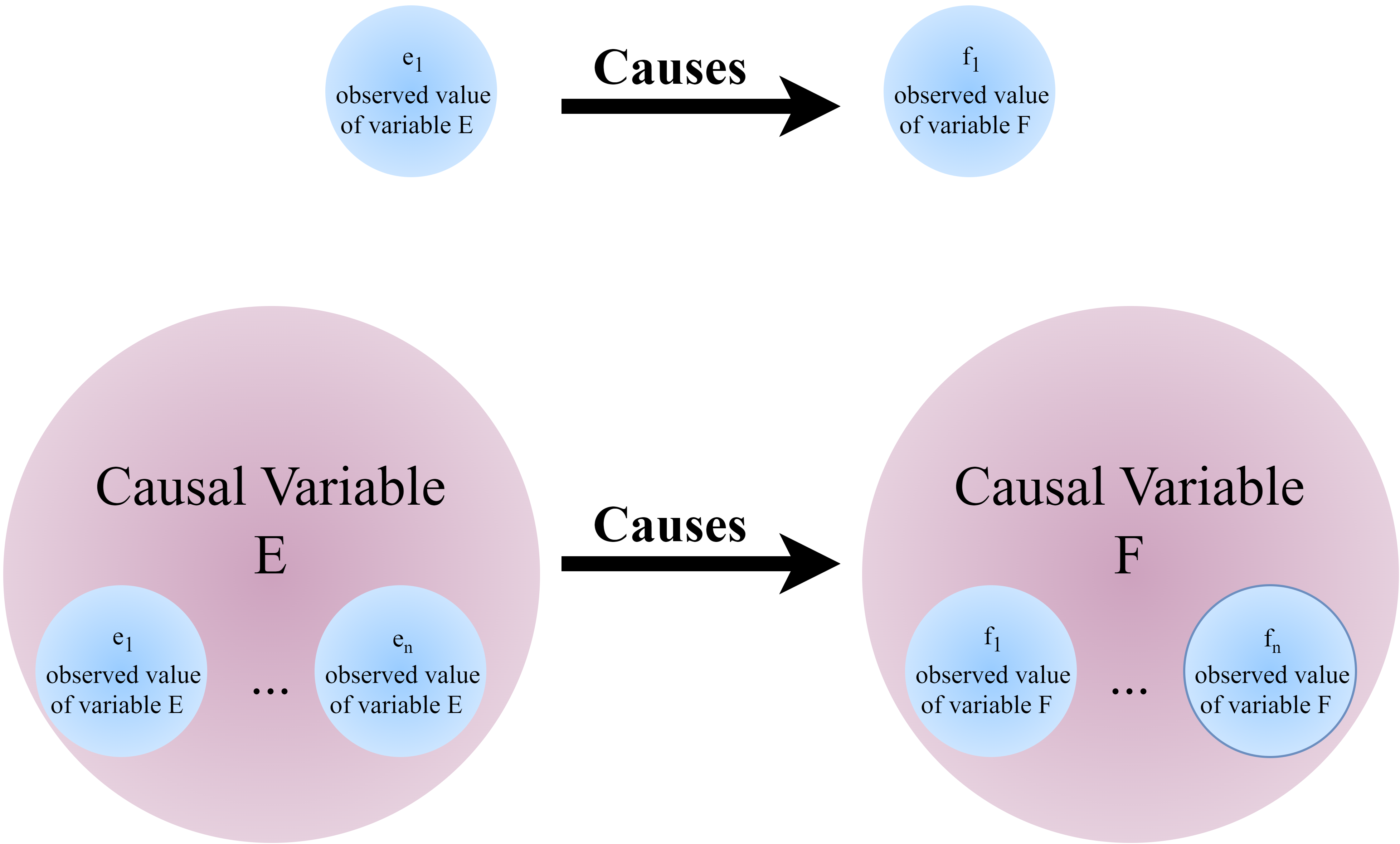}
  \caption{\textbf{Identifying causal relations in causal diagrams.} In a causal diagram, two variables represented as nodes are causally connected if there is at least one text entity/value belonging to the first variable that is the direct cause of an text entity/value belonging to the second variable. Causal relations should only be included between two variables if there is no other variable that fully mediates their relationship. Causal relations that include an interaction entity as a cause or an effect should be highlighted as a different relation type, as causal diagrams are intended to link two causal variables rather than a causal variable to a interaction entity  represented as an artificial node.}
  \label{fig:Causal modeling-Rule_3}
\end{figure}

In a causal diagram where the nodes correspond to causal variables or artificial nodes, it is sufficient to present a causal relation if one value of a node directly causes another value of a different node.
The direction of the causal relation is from the former node to the latter node. 
This principle is emphasized by Suzuki et al.~\cite{suzuki2020causal}. 
However, it is important to note that causal relations should only be included for direct effects, not distant effects. 
In other words, causal relations should only be included between two variables if there is no other variable that fully mediates their relationship.
This recommendation is also highlighted by Razouk and Kern~\cite{razouk2022improving}. 
In situations where artificial nodes representing interaction entities are integrated into the causal diagrams as part of the causal relation, it is crucial to emphasize that the nature of this relation should be distinct from other causal relations in the diagrams. 
Specifically, the relation type does not conform to a standard causal relation, as causal diagrams are intended to link two causal variables rather than a causal variable to text entities represented by artificial nodes.
Therefore, it is essential to clearly delineate this unique type of relation within the diagram to accurately represent the underlying causal structure.
Figure~\ref{fig:Causal modeling-Rule_3} provides a visual representation of this principle.


\subsection{The transitivity principles of causal relations should be preserved}
\label{CMR4}

\begin{figure}
  \centering
  \includegraphics[width=0.9\textwidth]{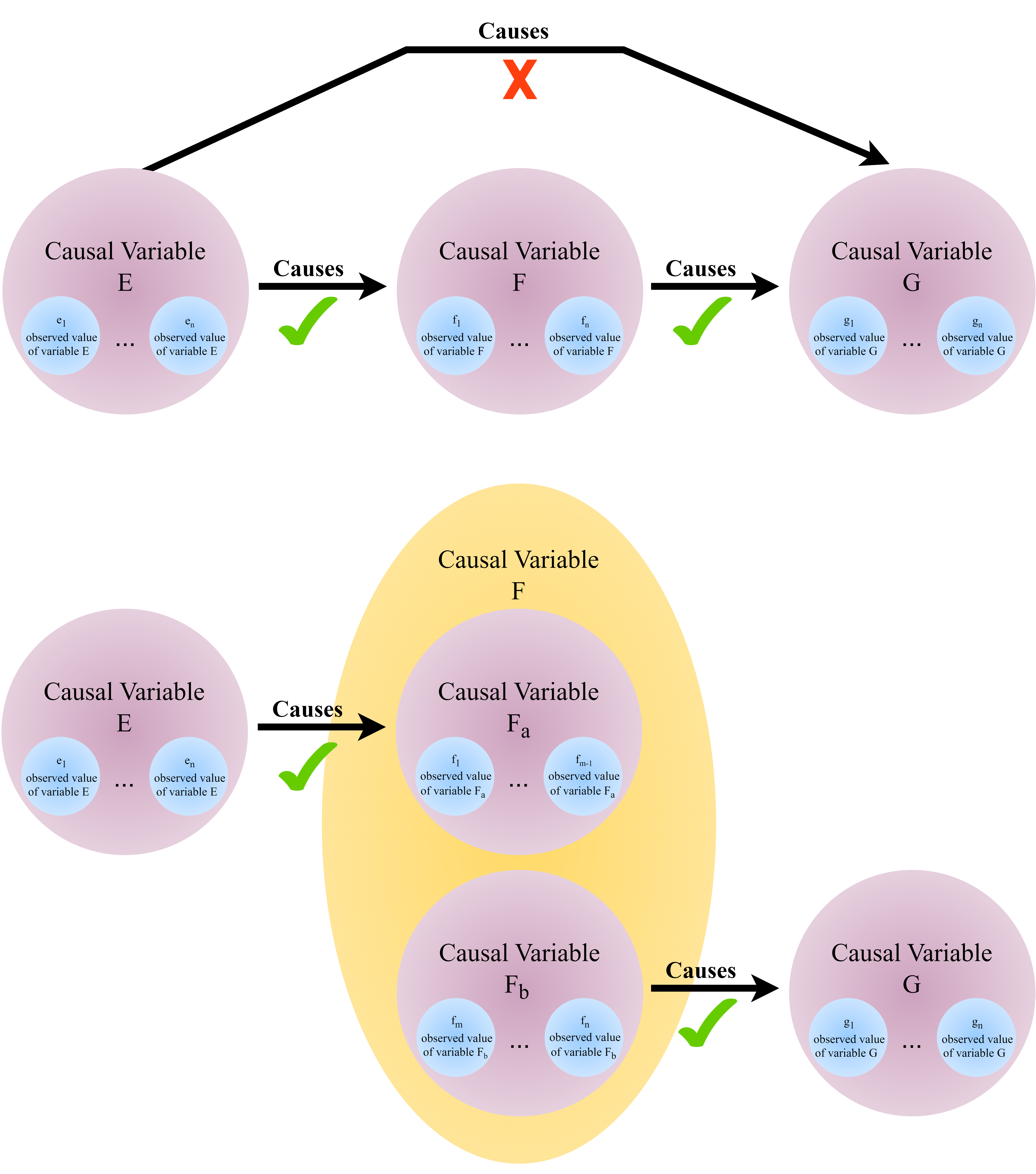}
  \caption{\textbf{Devising transitivity principles of causal relations to verify  causal variables encoding.}
    Violations to the transitivity principles of causal relations indicates impairments in the causal variables identifications. 
    One potential cause of this issue is that the identified mediator variable could be too general.
    Alternatively, at least a value of one of the mediator variable is actually an interaction even.}
  \label{fig:Causal modeling-Rule_4}
\end{figure}

In causal diagrams, the encoding of the causal variables can severely affect the validity of the modeled causal relations transitivity~\cite{mcdonnell2018transitivity}.
As such, when creating causal diagrams with causal paths involving more than two variables, it is crucial to ensure that the diagram accurately reflects the transitivity perception of the causal relation. 
To verify the validity of the identified causal variables, it is recommended to retrieve all the causal paths that have  three or more variables and check the validity of the transitive causal relation between the variables in the retrieved causal chains.
For instance, a causal path denoted as $CP$ describes two causal relations between causal variable $E$ and variable $F$ and between causal variable $F$ and $G$.
Each causal variable contains a set of values, like $f_1$, which denotes a value of causal variable $F$.
At least one of the values in causal variable $E$ causes one of the values in causal variable $F$. 
Similarly, at least one of the values of causal variable $F$ causes one of the values in causal variable $G$.
In case that the causal relation from $E$ to $G$ is not valid based on domain knowledge, this could indicate an issue in the identification of the causal variables.
One potential cause of this issue is that the identified mediator variable (i.e. in this example variable $F$) is too general and  contains values which describes different independent features which can be modeled as independent variables with the support of the recommendation provided in Section~\ref{CMR1}.
Alternatively, at least a value of one of the mediator variable is actually an interaction entity which should be modeled as artificial node with the support of recommendation provided in Section~\ref{CMR2}. 
In both cases, these variables need to be revised and modified accordingly.
Figure~\ref{fig:Causal modeling-Rule_4} illustrates a scenario where a mediator causal variable $F$ is too general and violates the transitive relation between causal variable $E$ and causal variable $G$.
To accurately model the causal relation in a causal diagram, it is recommended to split the causal variable $F$ into variables $F_a$ and $F_b$.
$F_a$ and $F_b$ are not causally connected as such there is no causal path between causal variable $E$ and causal variable $G$.


These four rules can be applied iteratively until the resulting causal models meet all the rules criteria. 
By applying these rules, the resulting cognitive map is more aligned with causal knowledge modeling. 
As such, the subsequent analysis results could be more reliable and less biased.

\section{Results}

Our research builds on the work presented in~\cite{pinto2023analyzing}. 
Specifically, we have digitized the cognitive map and the causal diagrams introduced in the aforementioned work.
The digitization process was conducted manually, involving the listing of all entries in the cognitive map and their corresponding clusters in the form of a table. 
Similarly, the provided causal diagrams were transformed into causal and effect relations and presented in table form, along with their corresponding clusters.
The digitized cognitive map is provided in Appendix Table~\ref{tab:Digitized cognitive map}, offering a comprehensive representation of the digitized cognitive map. 
Additionally, the digitized causal diagrams are included in Appendix Table~\ref{tab:Digitized causal diagrams}, providing a structured overview of the transformed causal diagrams.
Upon initial examination, it is evident that the clusters exhibit significant overlap in their entries. For instance, the entry "Nightlife" is common to both the Public Space and Social Context clusters. 
Similarly, "Parking Problems" is shared between the Public Space and Mobility clusters.
Furthermore, "Support for Youth Renting" is a shared entry between the Economic Context and Social Context clusters.

To address this observation, we have worked hard to reassign the described entries in the cognitive maps to their respective causal variables by applying the first rule from the suggested guidelines. 
Our findings indicate that this process has yielded successful outcomes in several instances.
For instance, the entries "little inspection" and "lack of inspection" have been identified as two values of the same variable, "Inspection". 
Additionally, the entries "Gas Leak Inside Buildings" and "Gas Leak Inside Dwellings" have been found to be similar.
Furthermore, entries such as "Accesses", "Lack of Accessibility", and "Good Accessibility" have been recognized as being part of the same variable, "Accessibility". 
Similarly, "Unoccupied Housing (Empties)" and "Vacant Buildings" are analogous, and thus can be grouped under the same causal variable, indicating the vacancy rate. 
Additionally, the similarity between "Arms Trafficking", "Drug Trafficking", "Crime" and "Street Criminality", some of which are present in the "Social Context" and  some of which are present in "Public Space" clusters, signifies a notable overlap between these clusters.
Moreover, based on our knowledge, "Requalification of Abandoned Housing Buildings" and "Requalification of Abandoned Business Buildings" are also similar and can be grouped accordingly.
In addition, "Noise" and "Quiet" are two entries that can be grouped together, as "Noise" and "Quiet" are mutually exclusive (i.e., the presence of Noise implies the absence of Quiet, and vice versa).

Following our exploration, we have focused on identifying interaction entities that provide insights into more than one causal variable. A notable example of such a case involves the text entity "Abandoned Housing" and "Housing Overcrowding".
"Abandoned Housing" refers to houses that are vacant and in a state of severe disrepair, while "Housing Overcrowding" pertains to houses that are not vacant and are occupied by an unsuitably large number of residents. 
These entries represent interaction entities as "Unoccupied Housing (Empties)" solely indicates the vacancy rate as a causal variable, without addressing the condition of the house (severe disrepair) or the suitability of the number of residents for the actual building in question.
To effectively model these text entities, we have applied the rule in Section~\ref{CMR2} and represented them as artificial nodes in the causal diagram.

Next of our analysis, we have identified the causal relations between the proposed causal variables and analyzed the relationships between text entities within these variables. 
This process involves examining the presence of cause and effect pairs, where the cause belongs to one variable and the effect belongs to another. 
This signifies the existence of a causal relation directed from the first variable to the second, as elaborated in Section~\ref{CMR3}.
As an example, consider the text entity "Lack of Urban Policies" within the causal variable "Urban Policies", which may lead to the text entity "Lack of Inspection" within the causal variable "Inspections".
This indicates a causal relation from the variable "Urban Policies" to the variable "Inspections", thereby insuring its inclusion in the causal diagram. 
This systematic approach of identifying and incorporating such causal relations enhances the comprehensive representation of the interdependencies between the variables in the causal diagram.

In the next step, we investigated the digitized causal diagrams presented in~\cite{pinto2023analyzing} to provide an example of the application of the rule outlined in Section~\ref{CMR4}. 
Upon review, it is observed that these causal diagrams contain several loops, and the nomenclature of the entries differs slightly from that of the cognitive map, leading to some confusion. 
Specifically, discrepancies exist between the lowercase entries in Table~\ref{tab:Digitized causal diagrams} and the original entries in Table~\ref{tab:Digitized cognitive map}. 
For instance, "Abandoned Housing" in the cognitive map is denoted as "abandoned dwellings" in the causal diagrams, while "Shady Places" in the cognitive map is referenced as "dark places" in the causal diagrams.

Nevertheless, when examining the causal path depicted as "Vacant Buildings" $Causes$ "abandoned dwellings" $Causes$ "Low Level or Insufficient Infrastructure" $Causes$ "Vacant Buildings", it becomes apparent that this causal path forms a loop.
However, upon considering the vacancy rate in the causal variable encompassing "Vacant Buildings", it is plausible that it may not have a direct causal relation with the level of infrastructure in the causal variable involving "Low Level or Insufficient Infrastructure".
This inconsistency suggests an encoding flaw in the causal model.
In this scenario, the encoding of "abandoned dwellings" as a causal variable is questionable, as it should instead be represented as an artificial node. 
Specifically, "abandoned dwellings" serves as an artificial node that signifies the building is vacant and in a state of great disrepair. 
While the vacancy of the building may not directly impact the level of infrastructure, the state of great disrepair could potentially influence the surrounding infrastructure.

\section{Discussion}

The initial examination of the digitized materials revealed significant overlap between clusters, indicating potential interdependencies and shared causal factors across different domains. 
This observation may indicate an impairments in the definition of the clusters and might results in biases due to their overlap.
This observation has prompted the reassignment of entries to their respective causal variables using the suggested rules. 
The successful application of these rules has led to the identification of shared causal variables and the grouping of similar entries, enhancing the clarity and accuracy of the cognitive maps.
Moreover, the identification of interaction entities has provided valuable insights into the complex interrelationships between variables, leading to the representation of these entities as artificial nodes in the causal diagram. 
This approach has added depth to the understanding of the causal relationships and highlighted the importance of accurately representing these relationships for a comprehensive analysis.

Upon examining the digitized causal diagrams, discrepancies in nomenclature and the existence of loops were uncovered, signaling potential encoding imperfections within the causal model. 
Notably, the disparity between the entries "Abandoned Housing" and "abandoned dwellings" is observed.
These variations introduce a level of confusion and may lead to challenges in aligning the entries across the cognitive map and the digitized causal diagrams.
Furthermore, a detailed examination of the integrity of the transitivity principle pertaining to the causal relations enabled the identification and resolution of certain loops within the causal diagram.
This realization emphasizes the significance of accurately representing causal variables and their interrelationships to ensure the integrity and reliability of the causal model.

Based on this analysis, the prevalent challenges and potential pitfalls associated with modeling causal domain knowledge using cognitive mapping and other brainstorming approaches in the context of urban blight analysis can be categorized into two main areas. 
Firstly, it was observed that many entries and clusters in the cognitive maps do not consistently represent causal variables. 
Specifically, clusters are overlapped, entries, in some cases, represent different values of the same causal variable, and entries, in other cases, represent interactions between multiple causal variables.
To address this, a transformation of these entries is necessary to align them more closely with the representation recommended in the field of causal data science. 
Guidelines outlined in Section~\ref{CMR1} and Section~\ref{CMR2} would support this transformation process.
Secondly, the proposed causal relations exhibit loops and numerous ambiguities as they are documented in~\cite{pinto2023analyzing}.
Resolving these ambiguities can benefit from the recommendations provided in Section~\ref{CMR3} and Section~\ref{CMR4}, offering valuable insights into refining the causal model.

\subsection*{Limitation}

It is important to acknowledge the limitations encountered during the analysis process, particularly with regard to the ambiguity surrounding certain entries and the restricted accessibility to the original experts involved in the study.
These constraints have highlighted the essential need for further endeavors aimed at addressing ambiguities and soliciting additional input to enhance the accuracy and reliability of the reassignment process.
Furthermore, it is essential to note that the provided four rules only encompass scenarios where temporal and spatial information have not been taken into consideration. 
Consequently, the presence of loops in the resulting causal diagrams arises. 
To mitigate such occurrences in future work, it should be possible to extend the provided rules to incorporate the concept of explicit variables introduced by Zhang et al. ~\cite{zhang2022causal}, explicitly encoding the temporal and spatial aspects of causal relations.

\subsection*{Future work}

The potential applications of these rules can be extended to broader contexts where brainstorming approaches are utilized to model causal domain knowledge. 
For instance, these rules can be applied to industrial risk assessment documents which typically employ brainstorming approaches~\cite{razouk2023improving}, which capture and document domain-specific causal domain knowledge.
These documents are lacking from a causal domain knowledge modeling prospective.
By addressing this gap, there is potential for better integration of the documented causal domain knowledge into downstream tasks such as causal inference.

\section{Conclusions}

Our work addresses the critical need to transform causal domain knowledge obtained through brainstorming approaches into a representation that aligns with the best practices of causal data science. 
By systematically defining causal modeling rules, we aim to empower practitioners in conducting more effective causal domain knowledge gathering and modeling through brainstorming approaches.
This initiative arises from the recognition of the challenges inherent in translating  domain knowledge into causal models, particularly in the context of urban blight analysis.

To achieve our objectives, we established four foundational rules for causal modeling. 
These rules are designed to facilitate the identification and modeling of causal variables sourced from a set of text entities articulated by domain experts.
Additionally, they encompass the identification of artificial nodes based on text entity understanding, the regulation of causal relations between variables, and the application of the transitivity principle to validate resulting causal diagrams and identify potential inconsistencies.
These rules form the backbone of our approach, ensuring a systematic and comprehensive transformation of domain knowledge into a structured causal representation.

As a practical application of our approach, we digitized an existing cognitive map developed for analyzing urban blight causes. 
Subsequently, we tested different rules for causal modeling directly on this digitized cognitive map.
The outcomes of this case study not only validate the effectiveness of our approach but also underscore the critical importance of collaboration between subject matter experts and experts in the filed of causal data science. 
This collaborative effort mitigates common pitfalls in causal modeling, highlighting the need for interdisciplinary cooperation in tackling complex challenges.

Our work marks an important step towards bridging the gap between domain knowledge gathering and causal modeling in the context of urban blight analysis. 
We envision that this integrated approach will not only yield more reliable results but also serve as a strong base for future research and practical applications in urban planning and community development.

\bibliographystyle{unsrt}  
\bibliography{references}  

\clearpage 
\vspace*{\fill} 
\begin{center}
\Large{\bfseries Appendix} 
\end{center}
\vspace*{\fill} 

\begin{table}
\footnotesize
\caption{\textbf{Tabular representation of the first 50 examples of the digitized cognitive map presented in~\cite{pinto2023analyzing}.} 
Several entries share between the clusters can be observed. 
Several entries can be grouped under one causal variable.}
\label{tab:Digitized cognitive map}
\begin{tabular}{p{2cm} p{7cm} p{6cm}}

    \hline 
    \\ 
    \textbf{Index} & \textbf{  Text entity } & \textbf{Cluster}  \\ \\  \hline
    \\  
    1 &     Blight & NA\\  
    2 &     Lack of Geographic Information of the Territory & Blight \\  
    3 &     Urbanism & Lack of Geographic Information of the Territory \\  
    4 &     Public Space & Lack of Geographic Information of the Territory \\  
    5 &     Social Context & Lack of Geographic Information of the Territory \\  
    6 &     Economic Context & Lack of Geographic Information of the Territory \\  
    7 &     Public Policy & Lack of Geographic Information of the Territory \\  
    8 &     Mobility & Lack of Geographic Information of the Territory \\  
    9 &     Vacant Buildings & Urbanism \\  
    10 &    Antiquity of the Building & Urbanism \\  
    11 &    Unoccupied Housing (Empties) & Urbanism \\  
    12 &    Abandoned Housing & Urbanism \\  
    13 &    Low Level or Insufficient Infrastructure & Urbanism \\  
    14 &    Bad Location of the Building & Urbanism \\  
    15 &    Adaptation of Buildings with Housing Quality & Urbanism \\  
    16 &    Construction of Buildings for People with Disabilities & Urbanism \\  
    17 &    Rehabilitation & Urbanism \\  
    18 &    Evictions of the Local Population & Urbanism \\  
    19 &    Risk of Defeat & Urbanism  \\  
    20 &    Earthquakes & Urbanism \\  
    21 &    Gas Leak Inside Buildings & Urbanism \\  
    22 &    Floods & Urbanism \\  
    23 &    Gas Leak Inside Dwellings & Urbanism \\  
    24 &    Fires & Urbanism \\  
    25 &    Lack of Schools & Urbanism \\  
    26 &    Existence of Expecting Land & Urbanism \\  
    27 &    Lack of Hospitals & Urbanism \\  
    28 &    Change in the Use of Buildings & Urbanism \\  
    29 &    Inherited Buildings & Urbanism \\  
    30 &    Illegal Buildings & Urbanism \\  
    31 &    Lack of Equipment & Urbanism \\  
    32 &    Lack of Services & Urbanism \\  
    33 &    Lack of Municipal Coercive Works & Urbanism \\  
    34 &    Construction in High Zones of the Territory& Urbanism \\  
    35 &    Requalification of Abandoned Housing Buildings & Urbanism \\  
    36 &    Requalification of Abandoned Business Buildings& Urbanism \\  
    37 &    Public Services & Public Space \\  
    38 &    Street Criminality & Public Space \\  
    39 &    Shady Places & Public Space \\  
    40 &    Unsanitary Places & Public Space \\  
    41 &    Public Lighting & Public Space \\  
    42 &    Lack of Policing & Public Space \\  
    43 &    Arrangements of Public Spaces & Public Space \\  
    44 &    Degraded Public Space & Public Space \\  
    45 &    Trash on Public Streets & Public Space \\  
    46 &    Lack of Green Spaces & Public Space \\  
    47 &    Noise & Public Space \\  
    48 &    Air Quality & Public Space \\  
    49 &    Nightlife & Public Space \& Social Context\\  
    50 &    Evictions of Monos & Public Space \\  
    \\  \hline 
\end{tabular}
\end{table}

\begin{table*}
\caption{\textbf{Tabular representation of three out of seven digitized causal diagrams presented in~\cite{pinto2023analyzing}.}
Discrepancies in nomenclature and the existence of loops can be observed.
}
\label{tab:Digitized causal diagrams}
\begin{tabular}{p{6cm} p{6cm} p{3cm}}

    \hline  
    \\ 
    \textbf{Cause} & \textbf{ Effect } & \textbf{cluster}  \\ \\  \hline 
    \\  
    Urbanism  & Urbanism  & Cluster of Clusters\\  
    Urbanism  &  Social Context & Cluster of Clusters \\  
    Urbanism  & Mobility  & Cluster of Clusters  \\  
    Public Spaces  &  Social Context  & Cluster of Clusters  \\  
    Economic Context  & Urbanism  &  Cluster of Clusters \\  
    Economic Context  &  Social Context  & Cluster of Clusters  \\  
    Economic Context  & Public Spaces  & Cluster of Clusters  \\  
    Economic Context  & Mobility  & Cluster of Clusters  \\  
    Mobility  &  Social Context  & Cluster of Clusters  \\  
    Mobility  & Public Spaces  & Cluster of Clusters  \\  
    Public Policy & Urbanism  & Cluster of Clusters  \\  
    Public Policy &  Social Context  & Cluster of Clusters  \\  
    Public Policy & Public Spaces  & Cluster of Clusters  \\  
    Public Policy & Economic Context  & Cluster of Clusters  \\  
    Public Policy & Mobility  &  Cluster of Clusters  \\  

    Vacant Buildings &abandoned dwellings & Urbanism \\  
    Vacant Buildings & renovations & Urbanism \\  
    abandoned dwellings & renovations & Urbanism \\  
    abandoned dwellings &Low Level or Insufficient Infrastructure & Urbanism \\  
    Low Level or Insufficient Infrastructure & Vacant Buildings & Urbanism \\  
    Low Level or Insufficient Infrastructure &abandoned dwellings & Urbanism \\  
    Low Level or Insufficient Infrastructure & renovations & Urbanism \\  
    Low Level or Insufficient Infrastructure & Evictions of the Local Population & Urbanism \\  
    Low Level or Insufficient Infrastructure &Illegal Buildings & Urbanism \\  
    Low Level or Insufficient Infrastructure &Lack of Equipment & Urbanism \\  
    Evictions of the Local Population & Vacant Buildings & Urbanism \\  
    Evictions of the Local Population &abandoned dwellings & Urbanism \\  
    Evictions of the Local Population & renovations & Urbanism \\  
    Evictions of the Local Population &Illegal Buildings & Urbanism \\  
    Illegal Buildings & Vacant Buildings & Urbanism \\  
    Illegal Buildings &abandoned dwellings & Urbanism \\  
    Illegal Buildings & Evictions of the Local Population & Urbanism \\  
    Illegal Buildings & renovations & Urbanism \\  
    Lack of Equipment & Vacant Buildings & Urbanism \\  
    Lack of Equipment &abandoned dwellings & Urbanism \\  
    Lack of Equipment &Low Level or Insufficient Infrastructure & Urbanism \\  
    Lack of Equipment & renovations & Urbanism \\   

    street crime & Degraded Public Space  & Public Spaces \\  
    dark places  &street crime & Public Spaces \\  
    dark places  & Degraded Public Space  & Public Spaces \\  
    placement of public spaces  &street crime & Public Spaces \\  
    placement of public spaces  & dark places  & Public Spaces \\  
    placement of public spaces  & Degraded Public Space  & Public Spaces \\  
    placement of public spaces  & Lack of Green Spaces  & Public Spaces \\  
    Degraded Public Space  &street crime & Public Spaces \\  
    Degraded Public Space  & dark places  & Public Spaces \\  
    Degraded Public Space  & Degraded Public Space  & Public Spaces \\  
    Lack of Green Spaces  & Degraded Public Space  & Public Spaces \\

    \\  \hline 

\end{tabular}
\end{table*}

\end{document}